\documentclass[a4paper,twoside]{article}

\usepackage{epsfig}
\usepackage{subcaption}
\usepackage{calc}
\usepackage{amssymb}
\usepackage{amstext}
\usepackage{amsmath}
\usepackage{amsthm}
\usepackage{multicol}
\usepackage{pslatex}
\usepackage{apalike}
\usepackage{hyperref}
\usepackage{enumerate}
\usepackage{multirow}
\usepackage{makecell}
\usepackage{comment}
\usepackage{lipsum}
\usepackage{SCITEPRESS}     
\newcommand{\specialcell}[2][c]

\begin{document}

\title{ERQA: Edge-Restoration Quality Assessment for Video Super-Resolution}

\author{\authorname{Anastasia Kirillova\sup{1}\orcidAuthor{0000-0002-0799-3135}, Eugene Lyapustin\sup{1}\orcidAuthor{0000-0002-2515-9478}, Anastasia Antsiferova\sup{1}\orcidAuthor{0000-0002-1272-5135} and Dmitry Vatolin\sup{1}\orcidAuthor{0000-0002-8893-9340}}
\affiliation{\sup{1}Lomonosov Moscow State University, Moscow, Russia}
\email{\{anastasia.kirillova, evgeny.lyapustin, aantsiferova, dmitriy \}@graphics.cs.msu.ru}}


\abstract{Despite the growing popularity of video super-resolution (VSR), there is still no good way to assess the quality of the restored details in upscaled frames. Some VSR methods may produce the wrong digit or an entirely different face. Whether a method’s results are trustworthy depends on how well it restores truthful details. Image super-resolution can use natural distributions to produce a high-resolution image that is only somewhat similar to the real one. VSR enables exploration of additional information in neighboring frames to restore details from the original scene. The ERQA metric, which we propose in this paper, aims to estimate a model’s ability to restore real details using VSR. On the assumption that edges are significant for detail and character recognition, we chose edge fidelity as the foundation for this metric. Experimental validation of our work is based on the MSU Video Super-Resolution Benchmark, which includes the most difficult patterns for detail restoration and verifies the fidelity of details from the original frame. Code for the proposed metric is publicly available at \url{https://github.com/msu-video-group/ERQA}.}

\onecolumn \maketitle \normalsize \setcounter{footnote}{0} \vfill

\section{\uppercase{INTRODUCTION}}
\label{sec:introduction}

As a fundamental image- and video-processing task, super-resolution remains a popular research topic. It has a wide range of applications, from low-complexity encoding\footnote{\url{https://www.lcevc.org/}} to old-film restoration and medical-image enhancement. Trends in quality assessment of upscaled videos and images are favoring estimation of statistical naturalness in combination with fidelity. But restoration fidelity is much more important than statistical naturalness for some tasks: small-object recognition (e.g., license-plate numbers) in CCTV recordings, text recognition, and medical-image reconstruction.

\begin{figure}[!ht]
\centering
{\epsfig{file = 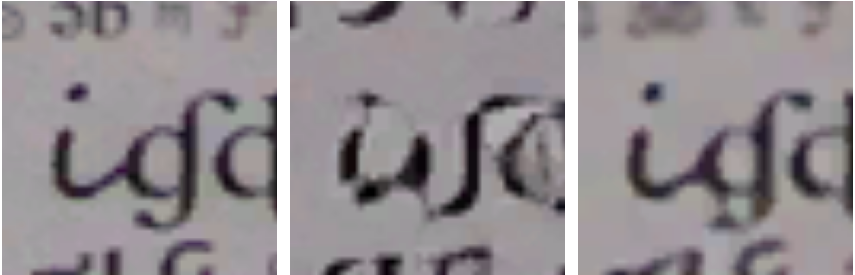, width = \linewidth}}
\caption{Example of changing context in an upscaled video: two characters from source frame (GT, leftmost) mix to yield a new one during video upscaling.}
\label{fig:artefact_1}
\end{figure}

\begin{figure}[!ht]
\vspace{-0.2cm}
\centering
{\epsfig{file = 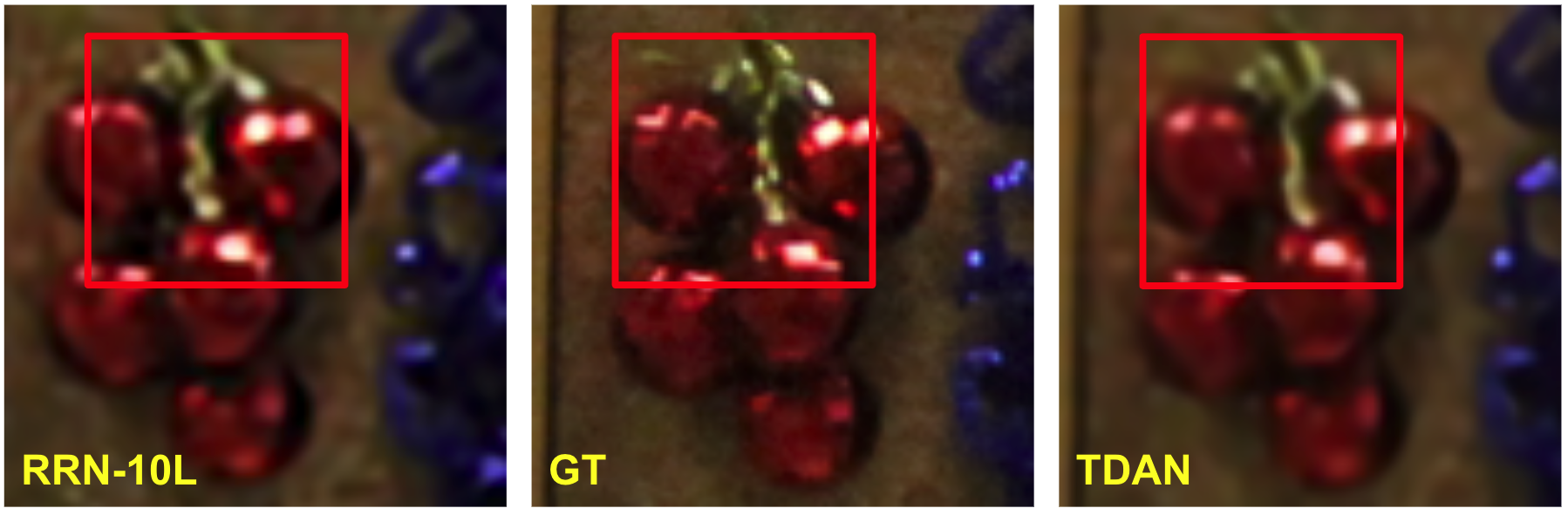, width = \linewidth}}
\caption{Example of upscaled images that vary in detail-restoration quality. The rightmost image is visually more natural, but the shape of the details in the leftmost image is closer to the original.}
\label{fig:artefact_2}
\vspace{-0.1cm}
\end{figure}

\begin{figure*}[!htb]
\centering
{\epsfig{file = 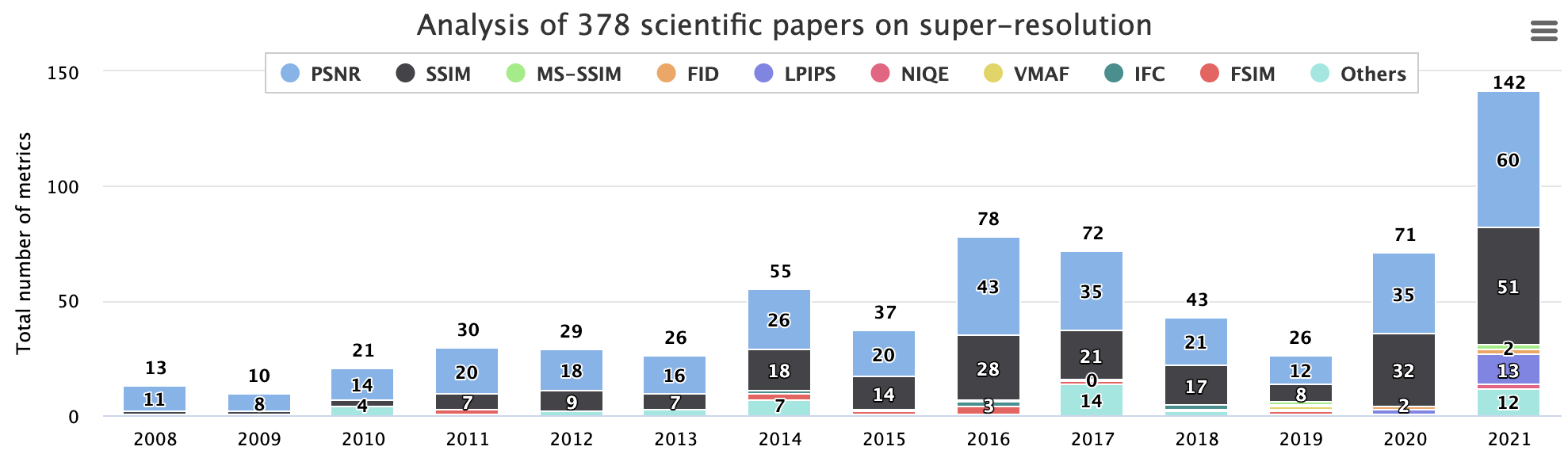, width = \linewidth}}
\caption{Metrics for estimating super-resolution quality cited in papers proposing new methods, by year. PSNR and SSIM \cite{wang2004image} are the most popular; LPIPS \cite{zhang2018unreasonable} saw wide use in 2020 and 2021.}
\label{fig:metrics_overview}
\end{figure*}

\begin{figure}[!ht]
\centering
{\epsfig{file = 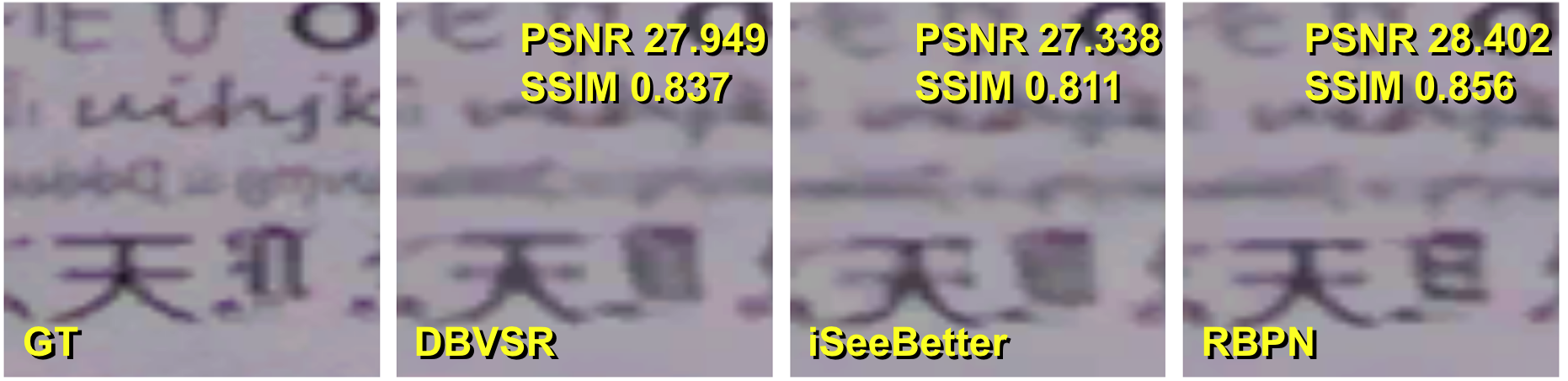, width = \linewidth}}
\caption{Example of changing context in an upscaled video: RBPN \cite{haris2019recurrent} has changed a character in the rightmost image.}
\label{fig:artefact_3}
\end{figure}
\begin{figure}[!ht]
\centering
{\epsfig{file = 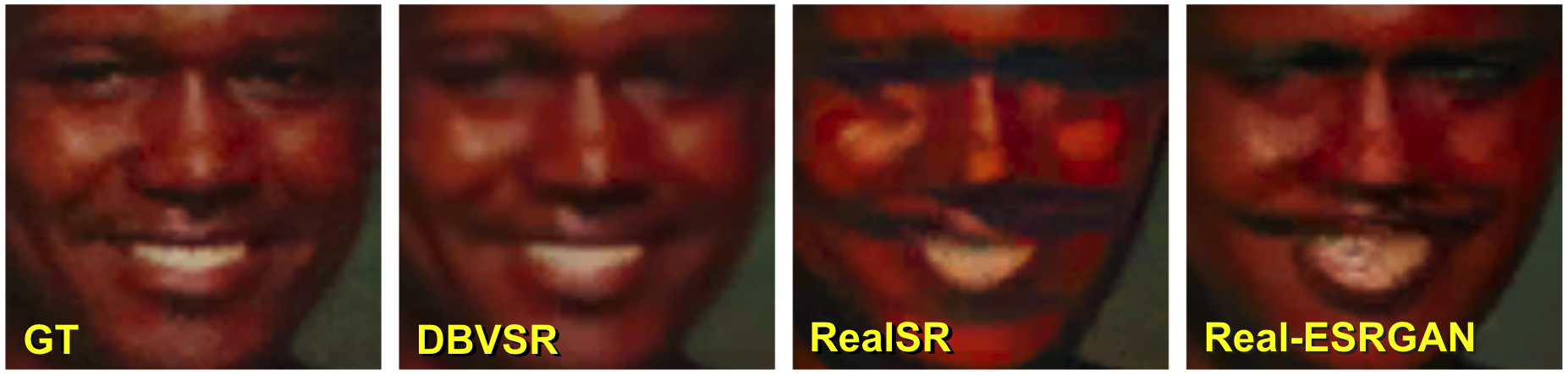, width = \linewidth}}
\caption{Example of changing context in an upscaled video: unnatural faces are the result here, differing considerably from the source one (GT, leftmost).}
\label{fig:artefact_4}
    \end{figure}
With the development of deep-learning-based approaches, many super-resolution models produce visually natural frames but lose important details. For example, the rightmost image in Figure~\ref{fig:artefact_2}, upscaled by TDAN \cite{tian2020tdan}, is perceptually better than the leftmost one, upscaled by RRN-10L \cite{isobe2020revisiting}, but the shape of the shiny thread in the leftmost image is closer to ground truth (GT, center). Occasionally, such models can even change the context in an image by, for example, producing an incorrect number, character, or even human face without decreasing traditional-metric values. In Figure~\ref{fig:artefact_1}, Real-ESRGAN \cite{wang2021real} mixed two letters from low-resolution images to form a completely different letter (center). In Figure~\ref{fig:artefact_3}, RPBN \cite{haris2019recurrent} added horizontal lines to the bottom-right character, but all three models score the same on traditional metrics. In Figure~\ref{fig:artefact_4}, Real-ESRGAN \cite{wang2021real} and RealSR \cite{ji2020real}  produced unnatural faces that greatly differ from the source one.

The examples in Figures \ref{fig:artefact_1}--\ref{fig:artefact_2}, \ref{fig:artefact_3}--\ref{fig:artefact_4} demonstrate that assessment of detail-restoration quality for image and video super-resolution is difficult. The best way to estimate restoration fidelity is to conduct a subjective comparison; it’s the most precise approach but is time consuming and expensive. Another way involves reference quality metrics. Traditional similarity metrics such as PSNR and SSIM \cite{wang2004image} are often used to evaluate super-resolution models, but they yield poor results and are unstable when dealing with shifts and other common super-resolution artifacts. LPIPS \cite{zhang2018unreasonable} is increasingly popular for this task, but it originally aimed to assess perceptual similarity rather than fidelity. The new DISTS \cite{ding2020image} metric is an improvement on LPIPS, but it also focuses on perceptual similarity.

Our research focuses on analyzing super-resolution algorithms, particularly their restoration fidelity. When we started working on a benchmark for video super-resolution\footnote{\url{https://videoprocessing.ai/benchmarks/video-super-resolution.html}}, including a test for restoration-quality assessment, we discovered that existing metrics work fine for other tests (restoration naturalness and beauty) but have a low correlation with subjective detail-quality estimation. In this paper, therefore, we introduce a new method for evaluating information fidelity. Experiments reveal that our metric outperforms other super-resolution quality metrics in assessing detail restoration.

\begin{table*}[!ht]
\caption{A comparison of datasets using for testing super-resolution quality assessment approaches.}\label{tab:datasets} \centering
\begin{tabular}{l c c c c}
  \hline
  Dataset & \# references & \# SR images & \# SR algorithms & Subjective type \\
  \hline
  \hline
  C. Ma et al.'s \cite{ma2017learning} & 30 & 1620 & 9 & MOS \\
  QADS \cite{zhou2019visual} & 20 & 980 & 21 & Pairwise comparison \\
  SupER \cite{kohler2019toward} & 14 & 3024 & 20 & Pairwise comparison \\
  SRIJ \cite{beron2020blind} & 32 & 608 & 7 & MOS \\
  SISRSet \cite{shi2019sisrset} & 15 & 360 & 8 & MOS \\
  ECCV \cite{yang2014single} & 10 & 540 & 6 & MOS \\
  SRID \cite{wang2017perceptual} & 20 & 480 & 8 & MOS \\
  \hline
\end{tabular}
\end{table*}
The main contributions of our work are the following:
\begin{enumerate}
\item A video-super-resolution benchmark based on a new dataset containing the most difficult patterns for detail restoration.
\item A subjective comparison examining the fidelity of details from the original frame, instead of traditional statistical naturalness and beauty.
\item A new metric for assessing the detail-restoration quality of video super-resolution.
\end{enumerate}

\section{\uppercase{RELATED WORK}}
\vspace{-0.6em}
PSNR and SSIM \cite{wang2004image} are common metrics for assessing super-resolution quality. We analyzed 378 papers that propose super-resolution methods and found that since 2008, PSNR and SSIM have remained the most popular metrics. But both have been shown to exhibit a low correlation with subjective scores. LPIPS \cite{zhang2018unreasonable} has grown in popularity over the last two years; other metrics remain less popular (Figure~\ref{fig:metrics_overview}).

Several full-reference metrics for assessing super-resolution visual quality have emerged. \cite{wan2018super} used four features (gradient magnitude, phase congruency, anisotropy, and directionality complexity) to calculate the perceptual structure measurement (PFSM) in both the upscaled and original high-resolution frames. Similarity function applied to PFSMs showed more-consistent results than previous approaches with regard to visual perception on their dataset. \cite{zhou2021image} calculated structural fidelity and statistical naturalness, fused these coefficients into a weighted sum, and achieved good correlation on the QADS image database\cite{zhou2019visual}.

Another popular approach is to extract structure or texture features from LR and upscaled (SR) images, compare them separately, and fuse the resulting similarity indices \cite{yeganeh2015objective,fang2019reduced}. Metrics based on this idea achieve a Spearman rank correlation coefficient (SRCC) coefficient of 0.69 to 0.85 on various datasets. \cite{yang2019machine} trained a regression model using statistical features extracted from LR and SR images, obtaining a correlation similar to that of other top metrics on the dataset from \cite{ma2017learning}. \cite{shi2019sisrset} proposed another approach for reduced-reference assessment that uses the visual-content-prediction model to measure the structure of the reference and SR images. This method outperforms previous ones on the SISRSet dataset \cite{shi2019sisrset}.

A number of no-reference metrics are also used for video super-resolution. \cite{ma2017learning} trains regression models on statistical features extracted from upscaled frames, achieving high value of SRCC on their dataset. \cite{zhang2021learning} proposed a no-reference metric, based on features extracted using a pretrained neural network—VGGNet. \cite{wang2018no} trained SVM using extracted features and obtained results similar to those of other metrics. \cite{greeshma2020super} proposed the SRQC metric, which estimates structure changes and quality-aware features. This metric exhibits good results, but they consider only a few images and four SR methods for the test dataset.

Edges have a strong influence on the human visual system. Furthermore, edge fidelity is a base criterion for assessing detail-restoration quality. Several methods thus consider edge features as the basis for quality assessment. Some calculate edge features, including number, length, direction, strength, contrast, and width, and compare them using the similarity measure to estimate image or video quality \cite{attar2016image,ni2017esim}. Nevertheless, these metrics achieve on their datasets almost the same correlation as traditional PSNR and SSIM. In \cite{xue2011image}, the authors detected edges in both reference and distorted images and compared them by calculating recall. \cite{chen2011image} used histogram analysis for edge comparison. These metrics deliver a slightly greater correlation than PSNR and SSIM. Liu et al. \cite{liu2019image} proposed using the F1 score to evaluate edge fidelity, but they declined to conduct a comparison with other metrics and kept their code under wraps. Our method is based on the same edge-comparison idea, but it’s robust for small local and global edge shifts, which appear during super-resolution but are unessential for detail recognition. It yielded much better results than other quality-assessment approaches.

A number of datasets are used for testing super-resolution quality assessment (Table~\ref{tab:datasets}), but not for detail restoration, because they lack difficult patterns for that task as in Figure~\ref{fig:test-stand} (text, numbers, QR codes, faces, complex textures). Therefore, we built a dataset for assessing super-resolution quality that includes the most challenging content for detail restoration.

Summarizing the above analysis, few metrics aim to assess and compare detail-restoration quality. Some that use edge features have emerged, but no one uses them for super-resolution, which involves peculiar artifacts. Therefore, it’s important to obtain an objective metric that correlates highly with human estimation of detail-restoration quality and that allows comparison of super-resolution models, not only for naturalness but also for information fidelity.

\section{\uppercase{PROPOSED METHOD}}
\subsection{Dataset}
\vspace{-0.6em}
To analyze a VSR model’s ability to restore real details, we built a test stand containing patterns that are difficult for video restoration (Figure~\ref{fig:test-stand}).

To calculate metrics for particular content types and to verify how a model works with different inputs, we divide each output frame into parts by detecting crosses:
\begin{enumerate}[{Part} 1]
\item ``Board'' includes a few small objects and photos of human faces\footnote{Photos were generated by \url{https://thispersondoesnotexist.com/}}. Our goal is to obtain results for the model operating on textures with small details. The striped fabric and balls of yarn may produce a Moire pattern (Figure~\ref{fig:moire}). Restoration of human faces is important for video surveillance.%
\item ``QR'' comprises multiple QR codes of differing sizes; the aim is to find the size of the smallest recognizable one in the model’s output frame. A low-resolution frame may blend QR-code patterns, so models may have difficulty restoring them.
\item ``Text'' includes two kinds: handwritten and typed. Packing all these difficult elements into the training dataset is a challenge, so they are each new to the model as it attempts to restore them.
\item ``Metal paper'' contains foil that was vigorously crumpled. It’s an interesting example because of the reflections, which change periodically between frames.
\item ``Color lines'' is a printed image with numerous thin color stripes. This image is difficult because thin lines of similar colors end up mixing in low-resolution frames.
\item ``License-plate numbers'' consists of a set of car license plates of varying sizes from different countries\footnote{The license-plate numbers are generated randomly and printed on paper.}. This content is important for video surveillance and dashcam development.
\item ``Noise'' includes difficult noise patterns. Models cannot restore real ground-truth noise, and each one produces a unique pattern.

\begin{figure}[!ht]
\centering
{\epsfig{file = 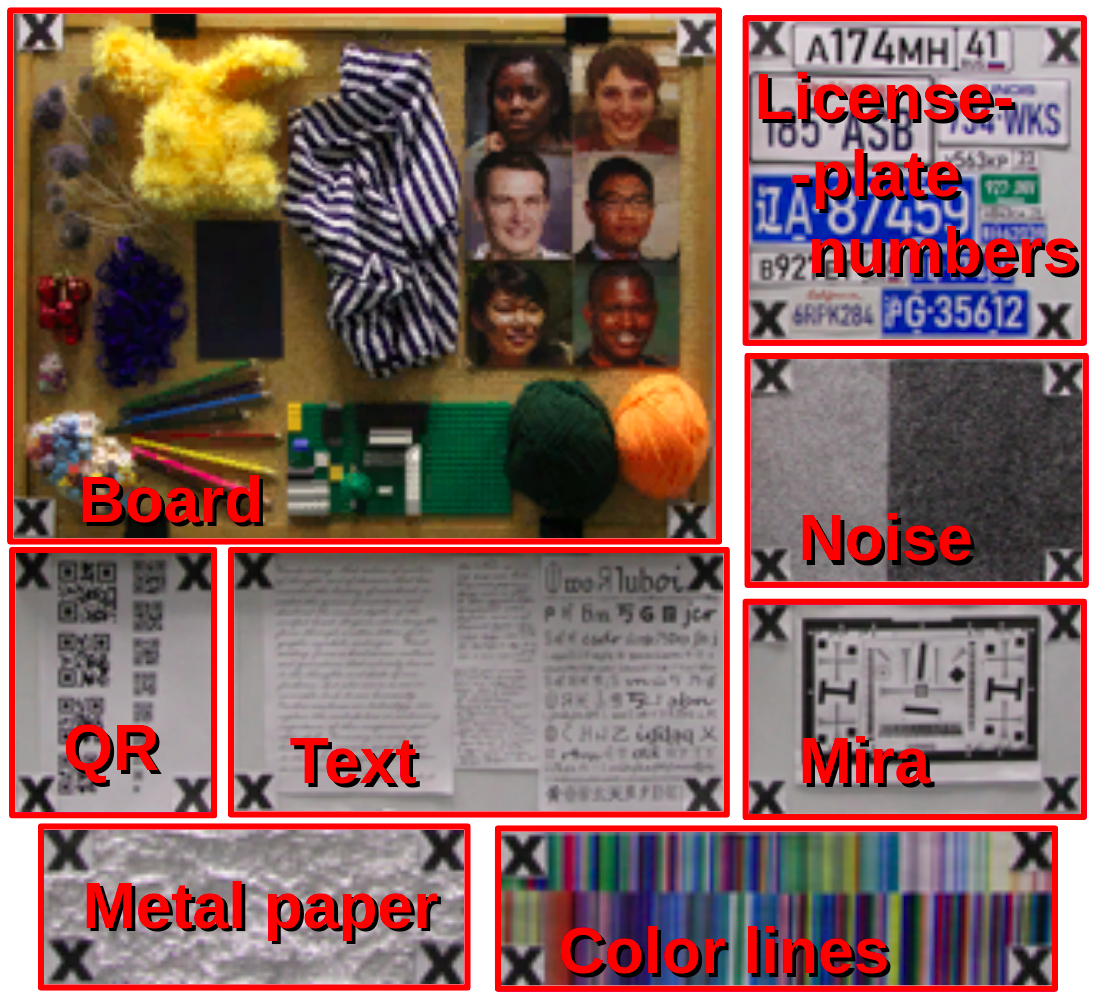, width = \linewidth}}
\caption{Test stand for the proposed VSR benchmark. The size of the stand $3,456\times3,456$ pixels in a source video and $1,280\times1,280$ pixels in a ground truth video.}
\label{fig:test-stand}
\end{figure}

\begin{figure}[!ht]
\centering
{\epsfig{file = 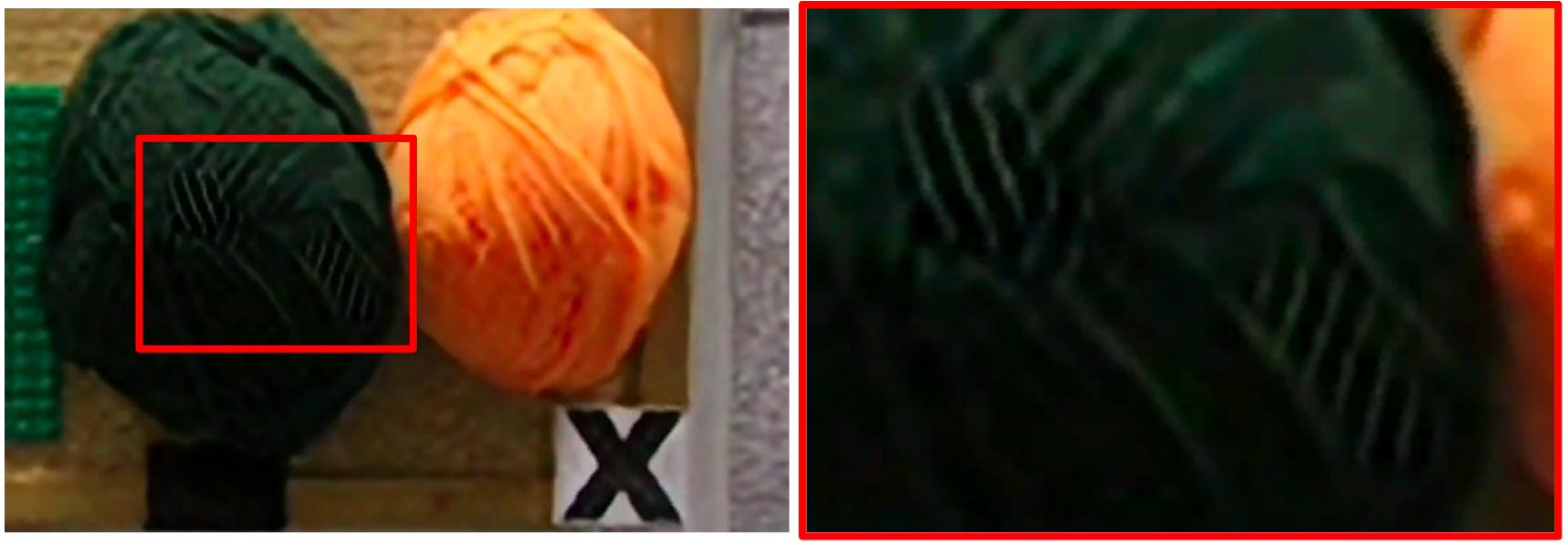, width = \linewidth}}
\caption{Example of a Moire pattern on the “Board.” There is a zoomed result on the right.}
\label{fig:moire}
\end{figure}

\item ``Mira'' contains a resolution test chart with patterns that are difficult to restore: a set of straight and curved lines of differing thicknesses and directions.
\end{enumerate}

We captured the dataset using a Canon EOS 7D camera. We quickly took a series of 100 photos and used them as a video sequence. The shots were from a fixed point without a tripod, so the video contains a small amount of random motion. We stored the video as a sequence of frames in PNG format, converted from JPG. The camera’s settings were ISO 4000, aperture 400, and resolution $5184\times3456$.

The source video also has a resolution of $5184\times3456$ and was stored in the sRGB color space. We degraded it using bicubic interpolation to generate a ground truth of resolution $1920\times1280$. This step is essential because many open-source models lack the code to process a large frame; processing large frames is also time consuming. We further degraded the input video from ground truth, again using bicubic interpolation, to $480\times320$ to test the models for $4\times$ upscaling. The output of each model is also a sequence of frames, which we compare with the ground-truth sequence to verify the model's performance.

\subsection{Subjective Comparison}
\vspace{-0.6em}
We used 21 super-resolution algorithms in our quality assessment. We also added a ground-truth video, so the experimental validation involves 22 videos. We cut the sequences to 30 frames and converted them to 8 frames per second (fps). This length allows subjects to easily consider details and decide which video is better. We then cropped from each video 10 snippets that cover the most difficult patterns for restoration and conducted a side-by-side pairwise subjective evaluation using the \url{Subjectify.us} service, which enables crowd-sourced comparisons.

To estimate information fidelity, we asked participants in the subjective comparison to avoid choosing the most beautiful video, but instead choose the one that shows better detail restoration. Participants are not experts in this field thus they do not have professional biases. Each participant was shown 25 paired videos and in each case had to choose the best video (``indistinguishable'' was also an option). Each pair of snippets was shown to 10-15 participants until confidence interval
stops changing. Three of pairs for each participant are for verification, so the final results exclude their answers. All other responses from 1400 successful participants are used to predict subjective scores using the Bradley-Terry  \cite{bradley1952rank}.

\subsection{Edge Restoration Quality Assessment Method}
\vspace{-0.6em}
On the basis of the hypothesis that edges are significant for detail restoration, we developed the edge-restoration quality assessment (ERQA) metric, which estimates how well a model can restore edges in a high-resolution frame. Our metric compensates for small global and local edge shifts, assuming they don’t complicate detail recognition.

\begin{figure}[!ht]
\centering
{\epsfig{file = 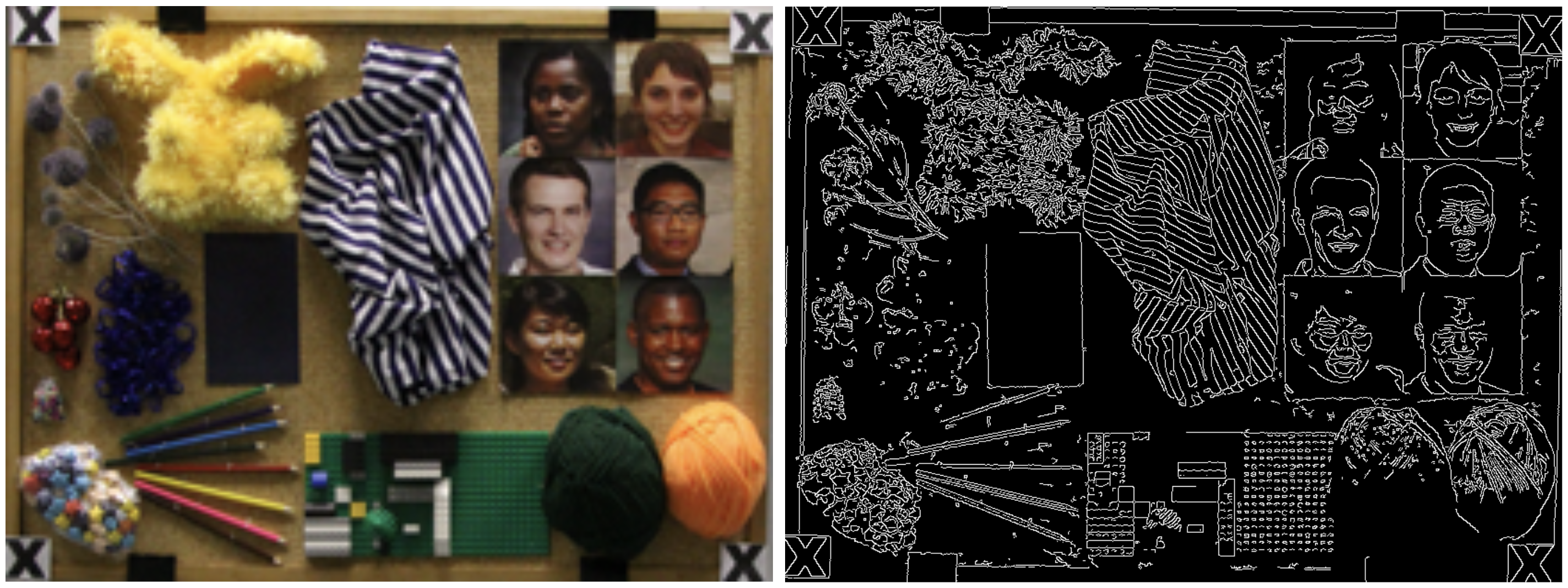, width = \linewidth}}
\caption{The content part "Board" cropped from a GT frame (left) along with edges of this frame highlighted with the Canny algorithm \cite{4767851} with chosen parameters (right).}
\label{fig:edges}
\end{figure}

\begin{figure}[!ht]
\centering
{\epsfig{file = 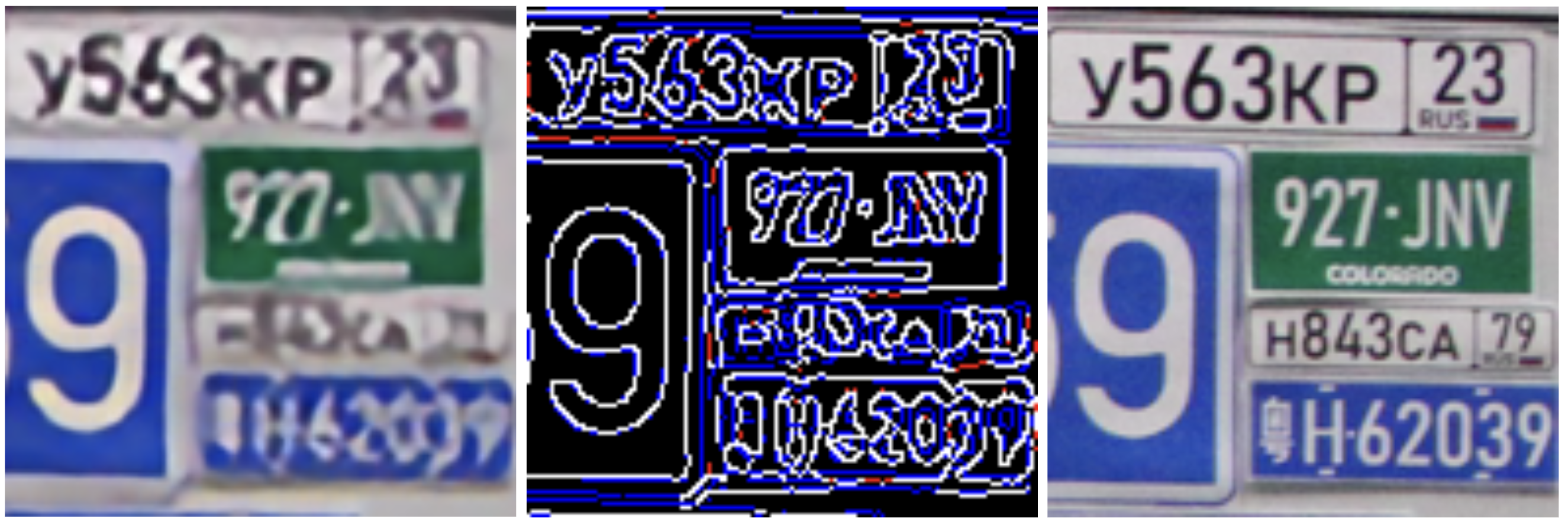, width = \linewidth}}
\caption{Crop from an upscaled frame (left), crop from the source frame (right) and visualization of ERQA metric (center). White = true positive, blue = false negative, red = false positive.}
\label{fig:erqa}
\end{figure}

First, we find edges in both the output and ground-truth frames. Our approach uses an OpenCV implementation\footnote{\url{https://docs.opencv.org/3.4/dd/d1a/group\_\_imgproc\_\_feature.html\#ga04723e007ed888ddf11d9ba04e2232de}} of the Canny algorithm \cite{4767851}. The threshold for initially identifying strong edges is 200, and the threshold for linking edges is 100. These coefficients allow us to highlight the edges of all objects, even small ones, while skipping lines, which are unimportant (Figure~\ref{fig:edges}).

Having found the edges in the ground-truth and distorted frames as binary masks, we compare them using the F1 score:

\begin{equation}\label{eq1}
    precision = \frac{TP}{TP + FP},\; recall = \frac{TP}{TP + FN},
\end{equation}

\begin{equation}\label{eq2}
    F_1 = 2 \frac{precision \cdot recall}{precision+recall},
\end{equation}

where TP (True Positive) is a number of pixels detected as edge in both ground-truth and distorted frames, FP (False Positive) is a number of pixels detected as edges only in distorted frame, FN (False Negative) is a number of pixels detected as edges only in ground-truth frame (Figure~\ref{fig:erqa}).

Some models can generate frames with a global pixel shift relative to ground truth, so we checked the integer pixel shifts $[-3,3]$ along both axes and chose the one with the maximum PSNR value. Compensating for this global shift aids our metric considerably (Table~\ref{tab:ablation_study}).

During an upscaling, models may also shift edge pixels locally, which in many cases is insignificant to human perception of information. To compensate for local single-pixel edge shifts, we consider as true positive any pixels on the output edges, which are not on the ground-truth edges but are near (on the difference of one pixel) with the edge of GT. 

We then noticed that some models produce a wider edge compared with the ground truth, and our method with local compensation (ERQAv1.0) marks these edges as fully true positive. To correct this shortcoming, ERQAv1.1 considers each point on a ground-truth edge as corresponding to true positive only once. The overall pipeline of Edge Restoration Quality Assessment method:

\begin{table}[htp]
\centering
\begin{tabular}{l l}
\hline
\multicolumn{2}{l}{Edge Restoration Quality Assessment pipeline} \\
\hline
\hline
\vspace{2mm}
Input:  & GT, image \\
Algorithm: & shifted\_img = global\_compensation \\
\multicolumn{2}{r}{(GT, image)} \\
 & GT\_edge = Canny(GT) \\
\vspace{2mm}
 & edge = Canny(shifted\_img)\\
 & TP, FP, FN = local\_compensation \\
\multicolumn{2}{r}{(GT\_edge, edge)} \\
Output: & ERQA = F1\_score(TP, FP, FN) \\
\hline
\end{tabular}
\end{table}

\begin{table*}[ht]
\caption{Spearman rank correlation coefficient (SRCC) of metrics with subjective assessment on all test cases. Metrics calculated on each test case compared with subjective score on the same test case. Mean value of correlation coefficients is also presented.}\label{tab:spearman_correlation} \centering
\begin{tabular}{l|c c c c c c c c c c|c}
  \hline
  Metric &   Lego & Toy &  Faces &  Yarn &    QRs & Text-1 &  Text-2 &  \shortstack{Car-1} &  \shortstack{Car-2} &   Mira &  \shortstack{Mean} \\
  \hline
  \hline
  ERQAv1.0           &  0.87 &            0.72 &   0.85 &           0.85 &  0.66 &               0.85 &          0.89 &           0.86 &           0.79 &  0.38 &             0.77 \\
  ERQAv1.1           &  0.87 &            0.66 &   0.89 &           0.84 &  0.65 &               0.85 &          0.91 &           0.92 &           0.88 &  0.41 &             0.79 \\
  SSIM*            &  0.68 &             0.20 &   0.81 &           0.33 &  0.52 &               0.57 &          0.63 &           0.86 &           0.86 &  0.29 &             0.58 \\
  PSNR*            &  0.36 &           -0.05 &   0.66 &           0.14 &   0.40 &                0.40 &          0.54 &           0.82 &           0.72 &  0.06 &             0.41 \\
  LPIPS              &   0.70 &            0.79 &   0.52 &           0.79 &  0.68 &               0.88 &          0.63 &           0.67 &           0.67 &  0.75 &             0.71 \\
  LPIPS*             &  0.78 &            0.81 &   0.56 &           0.84 &  0.69 &               0.88 &          0.65 &           0.72 &           0.71 &  0.75 &             0.74 \\
  DISTS              &   0.6 &            0.35 &   0.69 &           0.65 &  0.54 &               0.84 &          0.79 &           0.74 &           0.72 &  0.58 &              0.65 \\
  DISTS*             &   0.6 &            0.35 &   0.72 &           0.71 &  0.58 &               0.86 &          0.88 &           0.87 &           0.81 &  0.56 &             0.694 \\
  MS-SSIM           &  0.38 &             0.30 &   0.59 &            0.20 &  0.32 &               0.48 &          0.47 &           0.56 &           0.59 &  0.39 &             0.43 \\
  MS-SSIM*          &  0.77 &            0.19 &   0.68 &           0.35 &  0.48 &               0.53 &           0.6 &           0.82 &           0.81 &  0.25 &             0.55 \\
  VMAF  &  0.36 &            0.35 &   0.61 &           0.33 &  0.36 &               0.52 &          0.53 &           0.55 &            0.60 &  0.48 &             0.47 \\
  VMAF* &  0.33 &            0.36 &   0.57 &           0.38 &  0.34 &               0.52 &          0.48 &           0.56 &           0.58 &  0.47 &             0.46 \\
  VMAF (clip)              &  0.36 &            0.35 &    0.60 &           0.33 &  0.36 &               0.52 &          0.53 &           0.55 &           0.59 &  0.49 &             0.47 \\
  VMAF (clip)*              &  0.33 &            0.36 &   0.56 &           0.38 &  0.34 &               0.52 &          0.48 &           0.56 &           0.57 &  0.47 &             0.46 \\
  Ma et al.          &  0.47 &            0.88 &  -0.28 &           0.62 &  0.71 &                — &          — &            — &            — &   — &              0.48 \\
  \hline
\end{tabular}
\end{table*}

\begin{table*}[ht]
\caption{Performance comparison of all metrics with and without global compensation shifts.}\label{tab:global_shift} \centering
\begin{tabular}{|l|l l|l l|}
  \hline
   \multirow{2}{*}{Metric} & \multicolumn{2}{c|}{Without compensation} & \multicolumn{2}{c|}{With global pixel shift compensation} \\
   & PLCC & SRCC & PLCC & SRCC \\
  \hline
  \hline
  LPIPS & 0.8103 & 0.7077 & 0.8352 (+0.0249) & 0.7377 (+0.0300) \\
  DISTS & 0.8094	 & 0.6513 & 0.8278 (+0.0184) & 0.6931 (+0.0418) \\
  MS-SSIM & 0.2796 & 0.4282 & 0.5992 (+0.3196) & 0.5484 (+0.1202) \\
  VMAF & 0.2998 & 0.4692 & 0.2644 (-0.0354) & 0.4572 (-0.012) \\
  VMAF(not clipped) & 0.3428 &0.4706 & 0.2999 (-0.0429) & 0.4586 (-0.012) \\
  \hline
\end{tabular}
\end{table*}

\begin{table*}[!htb]
\caption{An ablation study of the proposed method.}\label{tab:ablation_study} \centering
\begin{tabular}{l|l l}
  \hline
  Stage & PLCC & SRCC \\
  \hline
  Without compensation (baseline) & 0.5035 & 0.4745 \\
  + Compensation of global shift & 0.7395 (+0.2360) & 0.6342 (+0.1597) \\
  + Compensation of local shift (v1.0) & 0.8243 (+0.0848) & 0.7383 (+0.1041) \\
  + Penalize false wide edges (v1.1) & 0.8316 (+0.0540) & 0.7519 (+0.0486) \\
  \hline
\end{tabular}
\end{table*}

\section{\uppercase{EXPERIMENTAL VALIDATION}}
\subsection{Ablation Study}
\vspace{-0.6em}
To verify the significance of the global- and local-shift compensation, we conducted a basic edge comparison without compensation, with only global compensation, with both global and local compensation (v1.0), and with penalization of wide edges (v1.1). All consistently increased both the PLCC and SRCC (Table~\ref{tab:ablation_study}).

We also tried our compensation scheme with Sobel, Robert\cite{roberts1963machine}, and Prewitt\cite{prewitt1970object} operators. Although there are some exceptions, in general ERQA shows a better correlation when using the Canny operator. Different thresholds for the Canny algorithm give metrics with high correlation with each other. Thus to avoid overfitting we empirically chose the theresholds 100 and 200 to highlight only important edges.

We also verified our metric on the QADS dataset \cite{zhou2019visual}. Although the mean correlation is lower than that of a few other metrics, the reason is that this dataset was developed for another test case (visual perception). In some situations, an image with lower visual perception looks more like the original one than does an image with higher visual perception. At the same time, working with images closer to our test-case ERQA yields good results.

\subsection{Comparison with Other Metrics}
We conducted a study of existing metrics for video-quality assessment and found that some work well for naturalness and beauty, but none works well for restoration. We calculated several well-known metrics on a new dataset: PSNR, SSIM \cite{wang2004image}, MS-SSIM \cite{wang2003multiscale}, VMAF\footnote{https://github.com/Netflix/vmaf}, the recently developed LPIPS \cite{zhang2018unreasonable}, which showed good results when assessing super-resolution imaging, its improvement DISTS \cite{ding2020iqa} and metric for SR assessment  \cite{ma2017learning}. Our metric outperforms all others in both the PLCC and SRCC. LPIPS places second. A popular metric for video-quality assessment, VMAF, exhibits poor results even compared with the traditional SSIM for this case. Multiscale structural similarity (MS-SSIM), which usually delivers better results than simple structural similarity (SSIM), ranked last on super-resolution.

We tried our global-shift-compensation scheme in an attempt to improve the performance of these metrics. Nearly all metrics (except VMAF) were better as a result (Table~\ref{tab:global_shift}).

Because metrics can work differently on different content types, we separately considered the correlation of metric values with subjective assessment on all crops and then calculated the mean correlation. Despite its simple and straightforward construction, ERQA delivers more-consistent results with subjective assessment and outperforms all other metrics in both the PLCC and SRCC (Table~\ref{tab:spearman_correlation}) coefficients when assessing information fidelity.

\section{\uppercase{Conclusion and Future Work}}
\vspace{-0.6em}
In this paper, we proposed a new full-reference ERQA metric for assessing detail restoration by video super-resolution. It compares edges in reference and target videos to analyze how well a VSR model restores the source structure and details. We also created a special dataset for assessing VSR quality and used it to analyze our metric through subjective comparisons. ERQA shows a high correlation with human detail perception and overall better results than traditional as well as state-of-the-art VQA methods. It approved that edge restoration are significant for human perception of detail restoration. The concept underlying our metric allows it to serve for similar restoration tasks, such as deblurring, deinterlacing, and denoising.

\section*{ACKNOWLEDGEMENTS}
\vspace{-1.4em}
Dataset preparation and subjective comparison were supported by Russian Foundation for Basic Research under Grant 19-01-00785 a. Metric development was supported by Foundation for Assistance to Small Innovative Enterprises under Grant UMNIK 16310GU/2021. Anastasia Antsiferova and Eugene Lyapustin were supported by the Fellowship from Non-commercial Foundation for the Advancement of Science and Education INTELLECT. 

%



\bibliographystyle{apalike}
{\small
\bibliography{example}}
\end{document}